\documentstyle[twocolumn,aps,epsf,prb]{revtex}

\begin{document}

\title{Sequencing of folding events in Go-like proteins}

\author{Trinh Xuan Hoang and Marek Cieplak}

\address{Institute of Physics, Polish Academy of Sciences,
Aleja Lotnikow 32/46, 02-668 Warsaw, Poland}

\address{
\medskip\em
{}~\\
\centering{
\begin{minipage}{14cm}
We have studied folding mechanisms of three small globular proteins:
crambin (CRN), chymotrypsin inhibitor 2 (CI2) and the fyn Src Homology
3 domain (SH3) which are modelled by a Go-like Hamiltonian with the
Lennard-Jones interactions.  It is shown that folding is dominated by
a well-defined sequencing of events as determined by establishment of
particular contacts.  The order of events depends primarily on the
geometry of the native state.  Variations in temperature, coupling
strengths and viscosity affect the sequencing scenarios to a rather
small extent.  The sequencing is strongly correlated with the distance
of the contacting aminoacids along the sequence.  Thus
$\alpha$-helices get established first.  Crambin is found to behave
like a single-route folder, whereas in CI2 and SH3 the folding
trajectories are more diversified. The folding scenarios for CI2 and
SH3 are consistent with experimental studies of their transition
states.
\end{minipage}
}}

\maketitle

\newpage 

\section{Introduction}

Go-like models \cite{Go} provide minimal yet fairly realistic
coarse-grained models of proteins. \cite{Takada,Baker} The main idea
is to attach importance only to those aminoacid-aminoacid interactions
which reside in the native contacts and then to choose the contact
energies that minimize the total energy in the native conformation.
This approach may seem to be overly simplistic but it generates models
with fast kinetics of folding.  In this sense, as noted by Takada
\cite{Takada}, "Go models may still be closer to reality than our
current realistic models". This is due to the fact that the presence
of significant non-native interactions can overconstraint and thus
frustrate a sequence of beads in continuum space \cite{Li1}.  Such a
structural frustration is expected to be of small consequence in fully
atomistic models that represent actual physical shapes of aminoacids.
One may then say that Go-like models appear to provide a mutual
compensation of two shortcomings in minimalistic models.  A useful
feature of the Go-like models is that they can be easily constructed
to describe realistic protein structures.  This ties well with the
finding that the geometry of of the native state itself has crucial
impact on the the foldability of proteins.
\cite{Baker,Plaxco,Wolynes,Micheletti,Maritan}

Lattice (see e.g. Ref.\cite{scaling}) and continuum space Go-like
models have been studied  in the literature.  There are several
versions of the continuum space Go models and they differ mostly in
what kind of the contact potential is used and in whether any steric
constraints are imposed or not.  Among the interactions that have been
considered there are: the square well potential,\cite{Zhou,Stanley}
the Gaussian function,\cite{Hardin} the Lennard-Jones (LJ) potential
\cite{Li1,Li,Hoang,Klimov2} and the short-range LJ-type potential with
the exponents of 12 and 10 in the repulsive and attractive parts
respectively.\cite{Clementi} It remains to be elucidated, however,
which of these functional forms is the most adequate.

In a recent paper,\cite{Hoang} we have reported results of a study of
several-sized $\alpha$-helices and $\beta$-sheets as modelled in the
Go fashion.  Our models employ the LJ potentials for the native
contacts and the soft repulsive potentials for the non-native
contacts.  The latter are necessary to provide excluded volume and
prevent entanglements. We have also considered models in which the
steric constraints are taken into account.  We have demonstrated that
in all of these models of secondary structures the folding proceeds
typically through well defined sequences of events which depend
primarily on the geometry of the native conformation.  This average
order of events is robust when the conditions for folding are optimal
but may become scrambled otherwise.  The helices are found to fold
preferably from the chain ends, whereas the $\beta$-hairpins usually
fold by starting from the turn.  Additionally, the formation of the
contacts has been found to proceed faster in the final stages of
folding.  Such a sequencing of contacts in the $\beta$-hairpin
formation resembles the kinetic zipping mechanism proposed in the
literature \cite{Dill,Munoz} and it agrees with recent simulations by
Klimov and Thirumalai \cite{Klimov2} and by Pande and
Rokhsar.\cite{Pande} The zipping mechanism has been found to dominate
also in a model of the $\beta$-hairpin that contain a hydrophobic core
(in which the interactions are stronger).\cite{Klimov2}
This indicates the important role played by the native geometry and it
validates results obtained based on the Go model.  Note, however, that
the all-atom simulations by Dinner et al.  \cite{Dinner} indicate
existence of pathways in which the hydrophobic core is established
first.

In the present study, we extend our Lennard-Jones based Go modelling
\cite{Hoang} to several small globular proteins which incorporate both
kinds of the secondary structures.  The proteins that we consider are:
a 46-monomer crambin (CRN), a 65-monomer chymotrypsin inhibitor 2
(CI2) and a 57-monomer Src Homology 3 domain of the fyn
tyrosine-protein kinase (SH3).  In order to simplify the analysis, we
model these systems without implementation of the steric constraints.
We demonstrate that our models form good folders.

Our main finding is that these models also generate a well defined
average order in which contacts are established provided the
temperature corresponds to optimal folding conditions.  The folding
history typically involves a steady establishment of successive
contacts which is interspersed by several characteristic temporal gaps
between certain stages.

Furthermore, we observe a good correlation between the separation of
the contacts along the sequence and the average times of their first
appearance during folding.  Long-ranged contacts usually need much
more time to get built than what the alpha helices need.  This finding
is consistent with recent experimental observation by Plaxco et al.
\cite{Plaxco} that the folding rates of proteins strongly depend on a
contact order parameter which is determined from the average sequence
separations between contacting residues in the native conformations.
The correlation between the time needed to  establish contacts and the
sequence separation is obeyed in CRN in an almost perfect fashion. In
CI2 and SH3, this correlation is disturbed but it remains strong.  The
idea that local interactions may dominate folding arose in studies of
simple lattice models.\cite{Unger} There are also, however, studies
which attach importance to the nonlocal contacts.\cite{Abkevich}

By examining the trajectories in details, we find that the typical
sequencing of folding events, as extracted from the average times of
establishing individual contacts, is the dominant folding scenario for
all of the systems considered.  In the case of crambin, for instance,
we find that 83\% of all trajectories follows the unique order of
events.  Such a high degree of determinism suggests that crambin may
be considered as the single-route folder.  For CI2 this degree of
determinism is lower but remains high: about 70\% of the trajectories
follow the typical folding scenario.  The typical sequencing of events
for CI2 also has been found to be in agreement with experimental
\cite{Jackson1,Jackson} and other simulation \cite{Clementi,Daggett}
results on the structure of the transition state.  The trajectories
for SH3 are found to be even more diversified with only 50\% being
typical.  However, in 75\% of the trajectories there is an early
establishment of the $3_{10}$-helix and the distal loop hairpin which
is consistent with experimental studies on the transition state.
\cite{Grantcharova,spectrin,Martinez1,Martinez,Riddle} We also find
that in 90\% of the trajectories the two $\beta$-sheets which are next
to the distal loop and the so called RT loop (defined in Figure 11)
are established earlier than the other segments.  This is in good
agreement with recent protein engineering analysis by Martinez and
Serrano, \cite{Martinez} which indicates that the folding of SH3 seems
to be composed of two folding subdomains.  

The sequencing of the folding events is found not to depend on the
viscous friction coefficient used in the simulations and moving the
temperature away from the optimal value results in a "scrambling"
which is similar to that found in isolated secondary structures.
Recently, there have been experimental studies
\cite{Jacob,Plaxco_friction} which reported that a large amount of the
solvent accessible surface area buried in the native state is also
buried in the transition state for a wide range of concentration of a
viscogenic agent.  Studies by Ladurner and Fersht \cite{Ladurner}
suggest that the viscogenic agent stabilizes the native state and the
transition state in parallel and to the same degree.  Thus the folding
mechanism appears not to depend on the solvent viscosity.  The folding
rate itself, on the other hand, has been reported to depend linearly
on the solvent viscosity for some proteins by the isostability
approach.\cite{Jacob,Plaxco_friction,Sosnick}

In most of the present paper, the amplitude of the LJ native
potentials is assumed to be fixed at a common value.  However, we have
also considered models in which certain contacts are made to be
substantially stronger. Such contacts correspond to the disulfide
bridges in crambin and to the hydrophobic contacts in the case of CI2
and SH3.  We find that the strengthening of the contacts affects the
time scales only marginally and the average sequencing order remains
unchanged.  This indicates that the sequencing of the contact
formation is determined by the geometry of the native conformation.
Our results are consistent with various experimental studies which
suggests that the folding transition states are conserved among
proteins which share the same overall native topology.
\cite{Martinez,Plaxco2,Chiti}

The paper is organized as follows. In Section 2, we present a short
description of the model and of the simulation method. In Section 3,
we discuss ways to delineate the native basin and we determine the
folding properties of the models.  The mechanisms of folding for each
of the proteins are discussed in Section 4.  Section 5 provides
conclusions.

\section{Models and Methods}

\begin{figure}
\epsfxsize=3.0in
\centerline{\epsffile{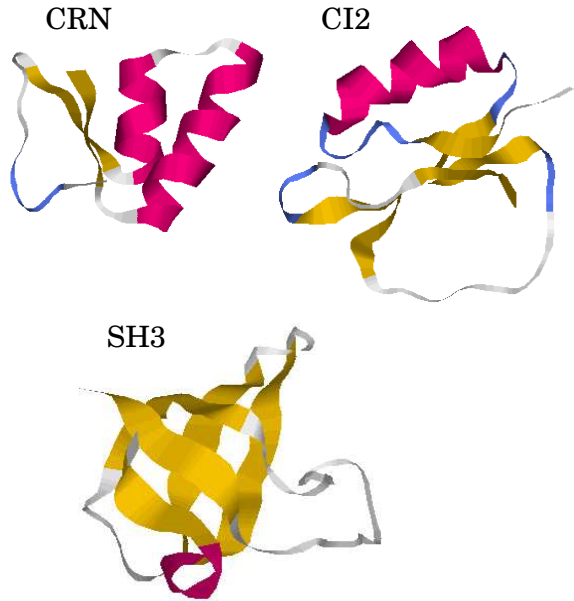}}
\caption{The native conformations of the proteins studied in the paper:
crambin (CRN), chymotrypsin inhibitor 2 (CI2) and the
SH3 domain (SH3). The PDB codes are 1crn, 2ci2 and 1efn.}
\end{figure}

We consider three single-domained globular proteins: CRN, CI2 and SH3.
The ribbon plots of their native conformations  are shown in Figure 1.
The native conformations involve at least one alpha helix and at least
two beta sheets in each case.

The proteins are modelled in a coarse-grained fashion, in which each
amino acid is represented by a single bead located at the position of
the C$\alpha$ atom. We adopt a simple Go-like interaction scheme and
do not implement any the steric constraints.

The brief summary of our approach \cite{Hoang} is as follows.
A chain conformation is defined by the set of position vectors
$\{ {\bf r}_i \}$, $i=1,2 \ldots N$, where $N$ is then number
of residues. The potential energy is assumed to take the 
form:
\begin{eqnarray}
E_p(\{{\bf r}_i\}) = 
\sum_{i=1}^{N-1} [ k_1 (r_{i,i+1} - d_0)^2 + k_2 (r_{i,i+1}-d_0)^4] \nonumber\\
+ \sum_{i+1<j}^{NAT}4\epsilon \left[ \left( \frac{\sigma_{ij}}{r_{ij}}
\right)^{12}-\left(\frac{\sigma_{ij}}{r_{ij}}\right)^6\right] \nonumber\\
+ \sum_{i+1<j}^{NON}4\epsilon \left[ \left( \frac{\sigma_0}{r_{ij}}
\right)^{12}-\left(\frac{\sigma_0}{r_{ij}}\right)^6 + \frac{1}{4}\right] 
\Delta(r_{ij}-d_{nat}). 
\end{eqnarray}
The first term in equation (1) represents rigidity of the backbone
potential; $r_{i,i+1}$ is the distance between two consecutive beads;
$d_0=3.8\AA$, $k_1=\epsilon$ and $k_2=100\epsilon$, where $\epsilon$
is the Lennard-Jones energy parameter corresponding to a native
contact.  The second term corresponds to interactions in the native
contacts and the sum is taken over all pairs of residues $i$ and $j$
which form the native contacts in the target conformation.  Two beads
that are not consecutive in the sequence are assumed to form a native
contact if their distance in the native conformation is less than
7.5$\AA$.  $r_{ij}$ is the distance between two residues $i$ and $j$,
and $\sigma_{ij} = 2^{-1/6}\cdot d_{ij}$, where $d_{ij}$ is the
corresponding native contact's length.  The third term represents the
excluded volume interactions between monomers in the non-native
contacts.  We chose $d_{nat} = \left<d_{ij}\right>$ and $\sigma_0 =
2^{-1/6}\cdot d_{nat}$.  $\Delta(r_{ij}-d_{nat})$ is a cut-off
function which is equal to 1 for $r_{ij} \le d_{nat}$ and 0 otherwise. 

The proteins are studied by using molecular dynamics (MD) simulations
in which the temperature control is accomplished through the
Langevin noise term so that the equations of motion read
\begin{equation}
m\ddot{{\bf r}} = -\gamma \dot{{\bf r}} + F_c + \Gamma \;\;.
\label{eqlang}
\end{equation}
Here, ${\bf r}$ is a generalized coordinate of a bead, $m$ is 
the monomer mass, $F_c=-\nabla_r E_p$ is the conformation force,
$\gamma$ is a friction coefficient and $\Gamma$ is the random force
which is introduced to balance the energy dissipation caused by
friction.  $\Gamma$ is assumed to be drawn from the Gaussian
distribution with the standard variance related to temperature, $T$, by
\begin{equation}
\left<\Gamma(0)\Gamma(t)\right> = 2\gamma k_B T \delta(t),
\label{eqgam}
\end{equation}
where $k_B$ is the Boltzmann constant, $t$ denotes time
and $\delta(t)$ is the Dirac delta function.

The equations of motion are integrated using the fifth order
predictor-corrector scheme,\cite{Allen} where the friction and random
force terms appear as a noise perturbing the motion at each
integration step.  The integration time steps are taken to be $\Delta
t = 0.005 \tau$ apart, where $\tau = \sqrt{m a^2/\epsilon}$ is a
characteristic oscillatory time unit. The characteristic length $a$ is
chosen to be equal to $5\AA$, which is a typical value of the Van der
Waals radius of the amino acid residues. The simulations are performed
with two values of the friction coefficient $\gamma$: $2m/\tau$, which
is a typical choice in the MD simulations of liquids, and $10m/\tau$
which we used before.\cite{Hoang} We have already demonstrated that
when $\gamma$ is larger than $1m/\tau$ the folding times scale
linearly with $\gamma$. Here, we show (in Section 4A) that the
ordering of the events is not affected by the value of $\gamma$, at
least in the case of CRN, so in the remaining study we stay with
$\gamma=2m/\tau$ to reduce the usage of the cpu. It should be noted
that the realistic $\gamma$ for amino acids in a water solution has
been argued \cite{Klimov} to be of order $50 m/\tau$.  In the
following, the temperature will be measured in the reduced units of
$\epsilon/k_B$.  

\section{Folding properties}

Proteins found in nature differ from random heteropolymers in that
their native states are not only thermodynamically stable but are also
easily accessible kinetically under the physiological conditions.  A
kinetic criterion \cite{Onuchic} for a sequence to be a good folder
(and thus to be a good model of a protein) is that the folding
temperature, $T_f$, which characterizes the thermodynamical stability,
is larger than the glass transition temperature $T_g$.  However, there
are problems with the definition of $T_g$: it depends on the value of
an arbitrary cutoff time and, more importantly, it presupposes
existence of the glassy phase.  A natural alternative is to use the
temperature, $T_{min}$,\cite{Coarse} at which folding is the fastest,
as a reference temperature. The bad folders are then those for which
$T_f$ is significantly lower than $T_{min}$.  An alternative
equilibrium criterion, on the other hand, specifies that good
foldability arises when the folding temperature $T_f$ is close to the
collapse transition temperature $T_\theta$.\cite{Camacho,Thirumalai}

In lattice models, the native state usually consists of just a single
microstate which simplifies determination of $T_f$ and of the folding
times.  $T_f$ is typically defined as a temperature at which the
probability of being in the native state crosses $\frac{1}{2}$. In the
off-lattice models, any conformation is of measure zero and the native
conformation has to be considered together with its immediate
neighborhood. A shape distortion method for calculation of the size of
a native  basin has been recently proposed by Li and Cieplak.\cite{Li}
The idea is to monitor the time dependence of the characteristic
conformational distance $\delta$ away from the native state based on
many short unfolding trajectories that evolve at different
temperatures.  $\delta$ is given by 
\begin{equation}
\delta^2=\frac{2}{N^2-3N+2}\sum_{i=1}^{N-2}\sum_{j=i+2}^N
(r_{ij}-r_{ij}^{NAT})^2,
\end{equation}
where $r_{ij}$ and $r_{ij}^{NAT}$ are the monomer to monomer distances
in the given conformation and in the native state
respectively. The distance $\delta$ is measured in $\AA$.
At a sufficiently large time scale the conformational
distance saturates below some critical temperature, $T_c$. The
saturation value of the distance at this temperature, $\delta _c$, is
used for the estimated native basin's size. 
Additionally, $T_c$ has been found \cite{Li,Hoang}
to be a measure of $T_f$. 

\begin{figure}
\epsfxsize=2.8in
\centerline{\epsffile{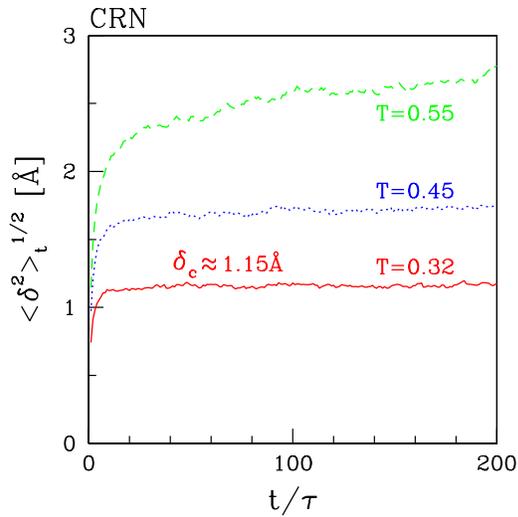}}
\caption{
The average root mean square
distance to the native state as a function of time for CRN.
The results for each temperature are averaged over 400
trajectories that start from the native conformation.
}
\end{figure}

Figure 2 illustrates the use of the shape distortion approach in the
case of CRN.  We considered 400 unfolding trajectories for each
temperature and computed the average root mean square distance
$\left<\delta^2\right>^{1/2}$ as function of time.  The estimated
basin size, $\delta_c = 1.15\AA$, is obtained at $T_c=0.32$ which is
the largest temperature at which the saturation is still observed. It
should be noted that, in the case of crambin, however, a precise
determination of $\delta _c$ and $T_c$ is not easy due to a broad
borderline behavior.  As shown in Figure 2, $T=0.45$ seems to be above
$T_c$  because there is a small but steady increase of $\left<\delta
^2\right>_t^{1/2}$ in time but the difference in the saturation or
near-saturation levels of the curves is rather  small.  Thus the value
of $\delta_c$ for CRN comes with a substantial error.  In the case of
CI2 (data not shown), the saturation is observed at $\delta_c$ of
about $4\AA$. Such a large value of $\delta_c$ does not seem to be a
realistic estimate of the basin size since it may correspond to not
all native contacts being established.  Thus the shape distortion
method appears not to be reliable when more complicated structures
than short chains are considered.  We then simply declare $1.15\AA$,
that we derived for crambin, to be a common approximate estimate of
$\delta _c$ for the three sequences studied here.

An alternative way to define the native basin is through the number of
the native contacts.  In the following we assume that two monomers
form a native contact if the distance between them is less than $1.5
\sigma_{ij}$.  Let $Q$ denote a fraction of the overlapping native
contacts between the native state and a given conformation.  $Q$=1
corresponds to a conformation that is in the native basin, whereas
$Q$=0 to a fully non-overlapping conformation.  It should be noted
that the contact criterion is not always equivalent to the criterion
which employs $\delta_c$.  For instance, we have checked that there
are conformations which have $\delta < \delta_c$ but $Q < 1$ and there
are also conformations which have $Q=1$ but $\delta > \delta_c$.  This
indicates a need for a more sophisticated multi-parameter description
of the native basin.

\begin{figure}
\epsfxsize=2.8in
\centerline{\epsffile{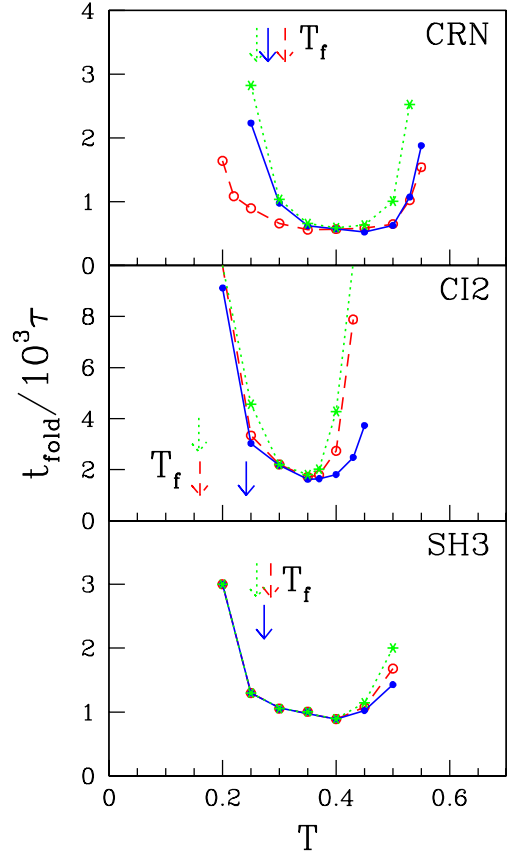}}
\caption{The temperature dependence of the median folding times
for the model proteins using three different criteria to define the native
basin: 1) $\delta < \delta_c$ (dashed line), 2) $Q$=1 (solid line)
and 3) both $\delta < \delta_c$ and $Q$=1 (dotted line), where $Q$ denotes
a fraction of the native contacts. The corresponding arrows indicate
positions of $T_f$ and $T_{min}$. }
\end{figure}

We have computed $T_f$ and the temperature dependence of the folding
times for the three model proteins studied here using three criteria
for what constitutes the native basin. These are:
1) $\delta < \delta_c$, 2) $Q$=1, and 
3) both $\delta < \delta_c$ and $Q$=1 which is the most
stringent criterion.
The median folding time at each $T$ are obtained based on typically
200 folding trajectories which start from random unfolded
conformations. The starting conformations are generated by performing
self-avoiding walks with randomly chosen bond angles and dihedral
angles. The bond angles are additionally restricted to be drawn form
the [$0$, $\pi/2$] interval in order to make the conformations be
shaped in an unfolded way.    
$T_f$'s are calculated based on 10 to 15 long $MD$ trajectories at
equilibrium at various temperatures. The trajectories start from the 
native state and last from $5\times10^4 \tau$ to $10^5 \tau$ depending 
on the system size. The first few thousands $\tau$'s are reserved
for equilibration and are not included in the averaging process.

Figure 3 shows that the median folding times follow the usual
$U$-shape dependence on temperature. For all cases, one observes that
the folding times at $T_{min}$ and $T_{min}$ itself do not depend on
the criteria used to define the native basin.  Away from $T_{min}$,
the three criteria work differently in each protein and the third
criterion yields the narrowest U shape.  A similar statement holds for
$T_f$ and the most stringent criterion yields the lowest value of
$T_f$.  Note, however, that overall the three criteria give similar
values of $T_f$ and in practice can be used interchangeably,
especially in the case of CRN and SH3.  In the case of CI2, the
difference between the $\delta_c$-based criterion and the
contact-based criterion is the largest.  We think that this large
difference is due to the presence of the active-site loop (shown in
Figure 10) which is poorly coupled to the rest of the structure.

\begin{figure}
\epsfxsize=3.2in
\centerline{\epsffile{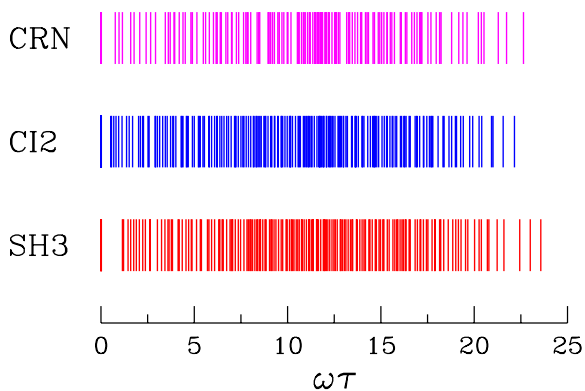}}
\caption{The phononic spectra of the studied proteins.
The values of $\omega_1 \tau$ given in radians are, top to bottom:
0.76, 0.51, 1.15}
\end{figure}

For the three model proteins studied here we observe that $T_f$'s are
comparable to $T_{min}$'s but are generally lower.  Note that at
temperatures around $T_f$ the native state can still be reached within
a time scale that is only about several times longer than the folding
time at $T_{min}$. Thus the sequences we study can still be considered
to be good or at least fair folders.  SH3 appears to be the best
folder of the three systems and CI2 to be the worst.  Note that if one
defines the glass transition temperature $T_g$ as one at which the
folding times become a few orders of magnitude longer (say, 1000
times) than the folding time at $T_{min}$, then for all of the three
models $T_f$ is larger than $T_g$.  

We have shown before \cite{Hoang} that phononic spectra for vibrations
around the native state offer some clues about the stability of
proteins: the bigger the gap in the low end frequency spectrum, the
bigger the mechanical stability, which should correlate with a bigger
value of $T_f$.  We repeat the calculation of the phononic spectra for
our model proteins. The procedure involves diagonalization of a $3N
\times 3N$ matrix of the elastic constants.\cite{Callaway} The elastic
constants are calculated numerically as the second derivatives of the
potential energy taken at the native conformation.  When calculating
an elastic constant for a bead we freeze all the remaining beads and
measure the resulting increase in the force when the unfrozen bead is
displaced along a given Cartesian direction.  The diagonalization is
done with the use of the Jacobi transformation method.\cite{recipes}
The resulting phononic spectra are shown in Figure 4.  The frequencies
$\omega$ are measured in units of $\tau^{-1}$.  In each spectrum there
are six zero frequency modes which correspond to the translational and
rotational degrees of freedom of the whole system.  The lowest
non-zero frequency $\omega_1$ (which  define the gaps in the spectra)
corresponds to the first excited mode.  For CRN, CI2 and SH3
$\omega_1$'s are found to be equal to 0.76$/\tau$, 0.51$/\tau$ and
1.15$/\tau$ respectively.  Notice that CI2 which has the lowest
$\omega_1$ is indeed the system with the lowest $T_f$.  For CRN and
SH3, however, the $T_f$'s are comparable but $\omega_1$ for CRN is
somewhat smaller than for SH3.  Thus the correlation between
$\omega_1$ and $T_f$ is not strong.  Note that $\omega _1$ which is a
direct measure of mechanical stability  which should provide bounds
against melting of the native conformation.  The stability measured by
$T_f$ is instead also a measure of the role of the non-native energy
valleys  --- can their presence reduce the probability of the sequence
staying in the native valley.

\section{Folding mechanism}  

We now discuss the order of folding events in the three model
proteins.  The simplified character of the model allows us to consider
hundreds of folding trajectories and to monitor individual contacts.
We focus on the native contacts and ask what are the characteristic
times $t_0$'s at which particular contacts are established if one
starts from an unfolded conformation.  In $\alpha$-helices and
$\beta$-structures the contacts were found to be getting established
in a well defined order \cite{Hoang} and we seek to find out if the
same holds in the case of larger sequences.

\subsection{Crambin (CRN)}

Crambin (CRN) is a small globular protein with only 46 amino acids.
Because of its small size, it has been a subject of many early MD
simulations (see e.g. Ref.\cite{Karplus}).  The native structure of
CRN exhibits a rich amount of the secondary structures.  There are two
$\alpha$-helices which are packed together with two short
$\beta$-strands, as shown in the bottom of Figure 4.  The structure is
additionally stabilized by three disulfide bridges (Cys3 - Cys40, Cys4
- Cys32 and Cys16 - Cys26) which improve the thermodynamic stability
significantly. 

\begin{figure}
\epsfxsize=2.8in
\centerline{\epsffile{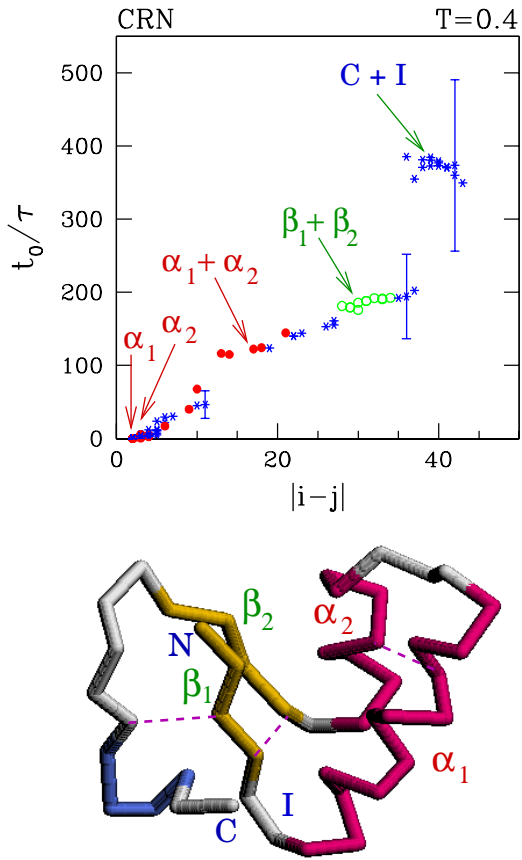}}
\caption{Top: the average times $t_0$'s needed
to establish contacts for the first time
during folding for CRN at $T$=$T_{min}$=0.4.
The times are plotted
against the sequence separations, $|i-j|$, between the monomers
that form the contacts.
The error bars indicate the
dispersions in the values of $t_0$'s. The error bars generally grow
with $t_0$ and several characteristic values are shown as an illustration.
The contacts are grouped into three categories: 1) those which belong to
the $\alpha$-helices (solid circles), 2) those which belong to
the $\beta$-sheets (open circles) and 3) other
(starred marks). Bottom: the native conformation of CRN in which
the various secondary structures are indicated.
N and C denote the N- and C-termini of the chain respectively. I denotes
a segment which connects the $\beta_1$-strand to the helix $\alpha_1$.
The dashed lines indicate the positions of the disulfide bridges.
}
\end{figure}

We first discuss the case of $\gamma=2 m/\tau$.  Figure 5 shows the
contact's establishment times $t_0$'s for CRN as calculated at
$T$=$T_{min}$=0.4.  The averages are taken over 400 independent
folding trajectories that start from random unfolded conformations.
The times are plotted against the sequence separation of the
contacting residues, denoted as $|i-j|$.  The error bars shown at
selected data points indicate the dispersions of these times.  One can
easily notice a strong correlation between the $t_0$'s and the values
of $|i-j|$.  The local contacts, i.e. the contacts with $|i-j|$ equal
to 2 or 3, are formed almost immediately.  The bigger the sequential
separation, the longer the average time that is needed for a contact
to start contributing to the energy. One can also observe a gap in
$t_0$'s between a group of contacts with the largest sequence
separations and the rest of the contacts. The big error bars shown for
some contacts indicate that the times and magnitudes of the gaps vary
substantially among the trajectories so it is simplest to start our
discussion by considering the average sequencing of events.

In order to understand the formation of the secondary structures
better we have divided the contacts into three groups: 1) the contacts
that are present in the $\alpha$-helices or link neighboring helices;
2) the contacts that are formed between the $\beta$-strands and 3)
other contacts, i.e. which belong to the coils, turns etc.

Figure 5 shows that contacts that belong to the helix group  are
formed the earliest, are all established within 150$\tau$, and remain
locked afterwards.  The contacts from the second group appear almost
simultaneously at about 200$\tau$, whereas the members of third group
become established at various time scales.  Notice that the tertiary
contacts between two helices are formed slower than within single
helices but faster than between the $\beta$-strands.  As shown in
Figure 5, the typical folding scenario is as follows.  First, the
$\alpha$-helices are established, then the two helices are locked
together.  Next, the contacts between the $\beta$-strands 1 and 2 are
established. At this stage, there is a long temporal gap and finally
the C-terminus gets connected to the rest of the structure which is
almost folded.

We now consider the case of $\gamma$ which is 5 times larger.  Figure
6a shows that the increase in $\gamma$ results in a five-fold
elongation of the characteristic times but the sequencing of the
contacts is not affected.  The pattern shown in Figure 6a is almost
identical to that shown in Figure 5.  This is in agreement with recent
experimental studies \cite{Jacob,Plaxco_friction,Ladurner,Sosnick}
which suggested that the folding mechanism does not seem to depend on
the solvent viscosity.

Figure 6b illustrates what happens if $T$ is changed from $T_{min}$ to
a lower value of 0.25.  The value of $\gamma$ is the same as in Figure
5: $2m/\tau$.  The change in the temperature increases the scatter in
the sequencing but the typical folding history remains unchanged.  A
similar observation holds also for temperatures which are higher than
$T_{min}$ (but are not too high).

\begin{figure}
\epsfxsize=2.8in
\centerline{\epsffile{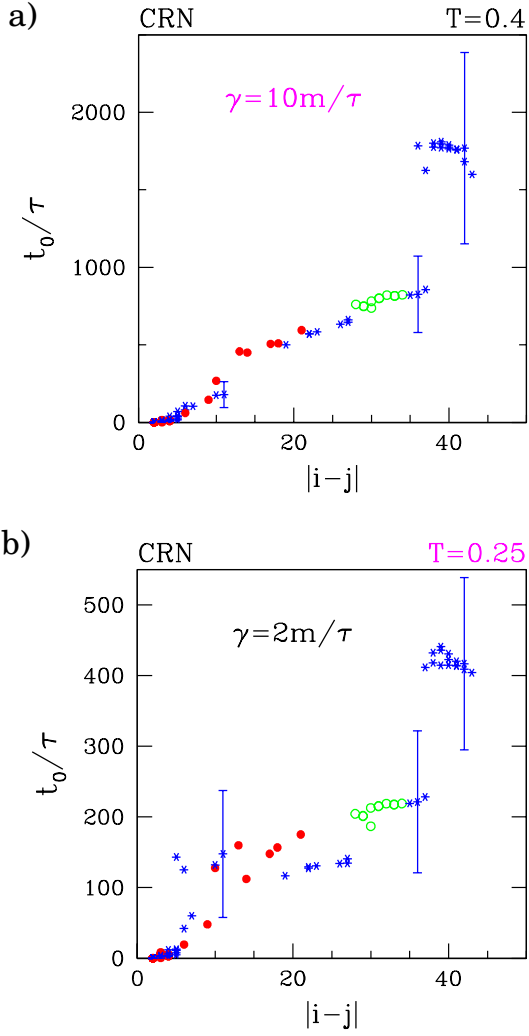}}
\caption{Same as the top of Figure 5, but
a) for $\gamma=10m/\tau$ and b) for $T=0.25$,
a temperature somewhat lower than $T_{min}$.}
\end{figure}

Results shown in Figures 5 and 6 have been obtained by averaging over
many trajectories.  Figure 7 shows a single typical folding trajectory
for CRN at $T$=0.4 and $\gamma=2m/\tau$, which exhibits a temporal gap
in the folding history as measured by the  contact establishment
times.  The top of the figure shows $t_0$'s and the bottom shows the
corresponding evolution in $Q$,  $E_p$, and the radius of gyration,
$R_g$.  The potential energy is seen to drift downhill and $R_g$ gets
shrinked.  $Q$ first increases monotonically but then it merely
oscillates for a period of about 300$\tau$ before entering the
vicinity of $Q$=1.  This period corresponds to the gap and is
indicated by the arrow.

\begin{figure}
\epsfxsize=3.2in
\centerline{\epsffile{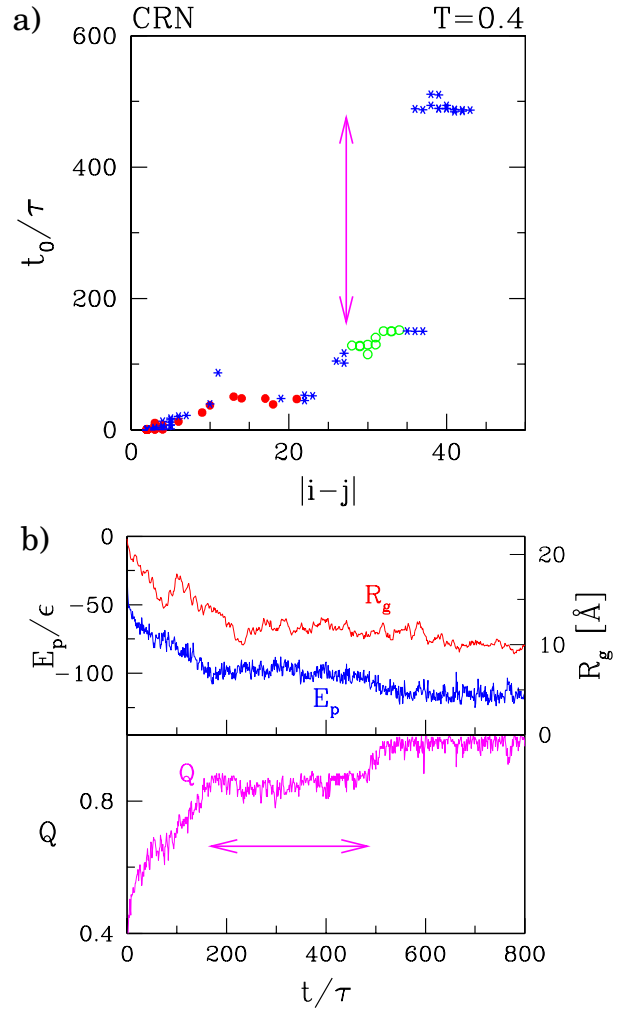}}
\caption{An example of a folding trajectory of CRN at $T=T_{min}=0.4$.
Top: the contact establishment times $t_0$'s versus the sequence
separation $|i-j|$. Bottom: the time dependence of the potential
energy $E_p$, the radius of gyration $R_g$, and the fraction of
the native contacts $Q$.
}
\end{figure}

Figure 8 shows selected conformations that appear on the trajectory
shown in Figure 7. It illustrates the folding scenario described in
Figure 5. The trajectory starts from a random unfolded conformation at
$t=0\tau$. At $t=80\tau$ one can observe the two helices are packed
together to form a bundle. Then the $\beta$-sheets appear as shown at
$t=160\tau$. At this moment there is one chain terminus that is still
far apart from the rest of the chain. It takes a significant amount of
time of about 330$\tau$ (corresponding to the gap in $t_0$'s) for this
terminus to move to its proper destination in the native conformation.
The last contacts start to form at $t\approx 490\tau$ and finally the
chain acquires its native shape, when all of the contacts are
established. This happens at $t=599\tau$.        

\begin{figure}
\epsfxsize=3.2in
\centerline{\epsffile{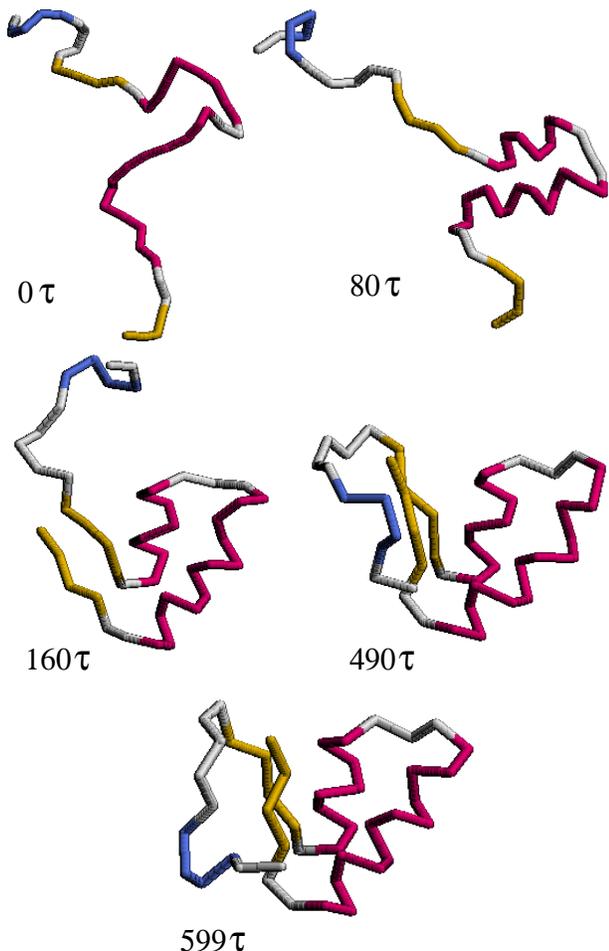}}
\caption{Examples of conformations probed from the trajectory shown
in Figure 7.}
\end{figure}

It should be noticed that the folding scenario presented above is just
the most likely scenario  and it follows from the geometry of the
native conformations. There can be other trajectories with a different
sequencing of events, depending on the initial conditions.  For
instance the C-terminus may be attached to the fragment I before the
$\beta_1$--$\beta_2$ sheet is formed.  In order to check the role of
the other scenarios we have examined 100 trajectories in details. We
find that among them there are:
\begin{itemize}
\item 83 trajectories with the scenario: \\
$\alpha_1 + \alpha_2 \longrightarrow \beta_1+\beta_2
\longrightarrow C+I$,
\item 9 trajectories with the scenario: \\
$\alpha_1 + \alpha_2 \longrightarrow C+I
\longrightarrow \beta_1+\beta_2$,
\item 5 trajectories with the scenario: \\
$C+I \longrightarrow \alpha_1 + \alpha_2
\longrightarrow \beta_1+\beta_2$, 
\item 3 trajectories with the scenario: \\
$\beta_1+\beta_2 \longrightarrow \alpha_1 + \alpha_2 
\longrightarrow C+I$.
\end{itemize}
We observe that the folding scenario presented by the average
sequencing of contacts is clearly dominating. 

In the discussion above all of the native contacts had the same
amplitude $\epsilon$ in the LJ potentials.  It is interesting to ask
what would happen if the amplitudes were not uniform. Specifically,
the interactions corresponding to the disulfide bridges are expected
to be significantly stronger and we ask if this could affect the
folding events.  There are three disulfide bridges in crambin,  as
indicated in Figure 5. Figure 9 shows the results on the event
sequencing when the interactions in the bridges is made five times
stronger.  One can notice that the characteristic times and the gap
become somewhat smaller but the essential physics of the sequencing
does not change. Thus the sequencing is primarily determined by the
topology of the native conformation.  

\begin{figure}
\epsfxsize=2.8in
\centerline{\epsffile{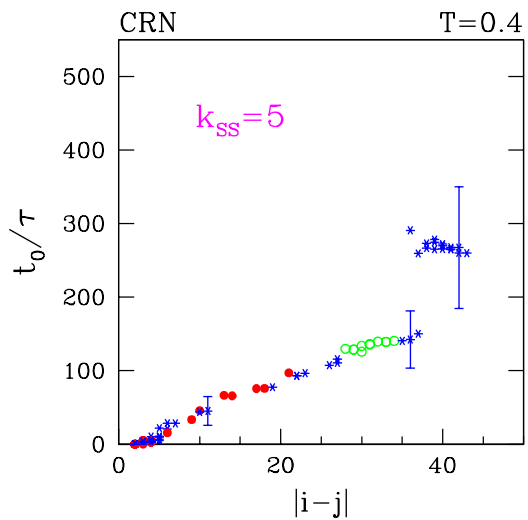}}
\caption{The sequencing of the contacts for CRN, at $T=0.4$
in the case in which the amplitudes of the potentials corresponding to
the disulfide bonds are five times larger than for the other
native contacts.}
\end{figure}

\subsection{Chymotrypsin Inhibitor 2 (CI2)}

In crambin, the order of events is perfectly correlated with the
sequence separation of the couplings. This is not so in the two other
systems although the degree of such correlations remains high. From
now on we work only with $\gamma =2m/\tau$.

We now consider CI2 which is a 83-residue protein.  The first 18
residues are not resolved by either x-ray crystallography or NMR, and
are considered to be irrelevant in protein engineering experiments
since they do not contribute to the stability or functionality of the
protein. The remaining 65 residues form a well-known native
conformation in which an $\alpha$-helix is packed against four
$\beta$-strands.  Experimental results show that CI2 folds rapidly by
a two-state mechanism.\cite{Jackson1,Jackson} 

In the present study, we examine the folding mechanism of CI2 along
the same lines as it has been done for CRN in the last section. Figure
10 shows the contact establishment times for CI2 obtained at its
temperature of the fastest folding ---  $T_{min}=0.35$. Similar to
CRN, there is a clear dependence of the average  $t_0$'s on the
sequence separation of the contacts, although this dependence is less
perfect.  The $\alpha$-helix is formed rapidly while the
$\beta$-sheets are established at various time scales.  The folding
scenario, on an average, can be presented as follows. After the
$\alpha$-helix is formed the $\beta$-sheet between two strands
$\beta_3$ and $\beta_4$ is also established.  Then, it takes about
800$\tau$ for the sheet between $\beta_2$ and $\beta_3$ to start to
form. Interactions between strands $\beta_1$ and $\beta_4$ appear at
the last stage of folding.  This happens just after several long range
contacts between the chain's segment connecting strand $\beta_1$ and
helix $\alpha$ (labeled by $I_1$ in Figure 10) and the coil fragment
connecting strands $\beta_3$ and $\beta_4$ (labeled by $I_2$) are
established.  There are two characteristic gaps in the time scales.
The first one is between the the $\beta_3$-$\beta_4$ interaction and
the $\beta_2$-$\beta_3$ interaction. The second one is between the
$\beta_2$-$\beta_3$ and the $\beta_1$-$\beta_4$ interaction.

\begin{figure}
\epsfxsize=3.0in
\centerline{\epsffile{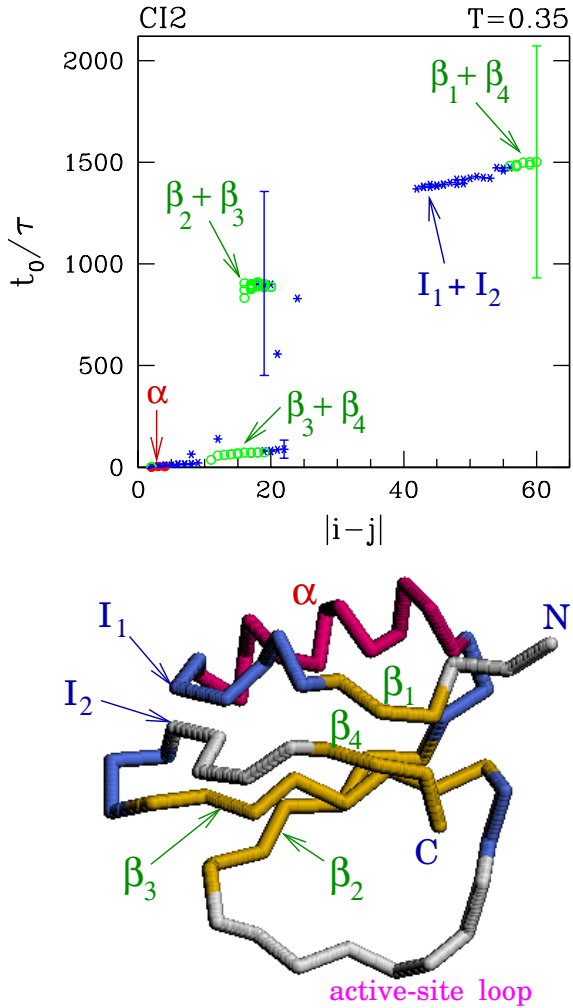}}
\caption{The same like Figure 5, but for CI2, at $T$=0.35.}
\end{figure}

Then we examine 100 trajectories and find that there are:
\begin{itemize}
\item 69 trajectories with the scenario: \\
$\beta_3+\beta_4 \longrightarrow \beta_2+\beta_3
\longrightarrow \beta_1+\beta_4$,
\item 15 trajectories with the scenario: \\
$\beta_2+\beta_3 \longrightarrow \beta_3+\beta_4 
\longrightarrow \beta_1+\beta_4$,
\item 15 trajectories with the scenario: \\
$\beta_3+\beta_4 \longrightarrow \beta_1+\beta_4
\longrightarrow \beta_2+\beta_3$,
\item 1 trajectories with the scenario: \\
$\beta_1+\beta_4 \longrightarrow \beta_3+\beta_4
\longrightarrow \beta_2+\beta_3$.
\end{itemize} 
Notice that for CI2 the trajectories are more diversified than
in the case of CRN  --- the CI2 is a larger system.
The folding scenario  corresponding to the
average sequencing of contacts still dominates.

Experimental \cite{Jackson} studies on CI2 have indicated that the
transition state ensemble of CI2 is broad. It means that there exist
many pathways to the native state and none of them is dominant, which
supports a ``new view'' of the protein folding.\cite{newview} Our
present analysis shows that there is a dominant sequencing of events
among the pathways --- at least when the events are measured in terms
of what contacts are established.  Note, however, that a given
sequencing may still correspond to multiple pathways.  The sequencing
of events found in CI2 is less pronounced compared to CRN. Thus, our
results are not in disagreement with the experimental results.
Studies on the structure of the transition state
\cite{Clementi,Jackson,Daggett} have indicated that in the transition
state the $\beta_1$-strand and the $\alpha$-helix interact weakly with
the rest of the structure. This observation also seems to be
consistent with our calculations on the sequencing.  As it has been
shown, the last folding event usually involves establishing the
contacts between the part of the protein which spans from the
N-terminus to the $\alpha$-helix and the rest of the structure.  This
latest rearrangement is also found to be  consistent with the
unfolding simulations for the full-atom model of CI2 \cite{Daggett} if
we assume that that unfolding corresponds to a reversed sequencing of
the events. 

In order to examine the influence of the hydrophobic core
on the sequencing of the events, we have  studied what happens
when strengths of 8 contacts between the hydrophobic residues
(Val28 - Ile76, Val28 - Val79, Val32 - Ile76, 
Ile48 - Val66, Val50 - Leu68, Val66 - Val82, 
Val70 - Ile76, Ile76 - Val79) are made stronger by the factor of 5.
We find that the average sequencing is affected very
little. The changes are even smaller than 
in the case of CRN when the disulfide bonds are 
made stronger.

\subsection{SH3 domain (SH3)}

There are a number of the Src Homology 3 (SH3) domains which all fold
to the same native conformation by a two-state mechanism:
src,\cite{Grantcharova}
$\alpha$-spectrin,\cite{spectrin,Martinez1,Martinez} fyn tyrosine
kinase \cite{Fyn} and phosphatidylinositol 3 -kinase
(PI3-kinase).\cite{PI3} Since the difference between the native
conformations of these domains are small, we will concentrate on the
SH3 domain of the fyn tyrosine kinase (PDB code: 1efn; residues
85-141).  The structure of this 57-residue peptide fragment is shown
in Figure 1 and 11 (bottom).  It has one small $3_{10}$-helix and five
$\beta$-strands which are packed together to form a $\beta$-sandwich.
The $\Phi$-value analysis of the fyn SH3 domain has been reported
recently,\cite{Plaxco2} after it was done for the two other
homologies: src \cite{Grantcharova} and
$\alpha$-spectrin.\cite{spectrin,Martinez1} In contrast to CI2, the
transition state of SH3 is found to be highly polarized.  This term
means that there exists a dominant route to the native state which has
a significantly lower free energy than all other routes.

The sequencing of the contacts for SH3 at $T_{min}$ of 0.4 is
presented in Figure 11.  One can observe a much weaker dependence of
$t_0$'s on the sequence separation of the contacts than in the case of
CRN and CI2.  The most distant contacts are established much earlier
than some middle range tertiary contacts.  The most possible folding
scenario is found to be as follows.  After the  $3_10$-helix is
established, the $\beta$-hairpin of the distal loop is also formed
immediately afterwards. At the same time, the contacts within the RT
loop start to become established which leads to the formation of the
sheet between the $\beta_1$ and $\beta_2$ strands. These above
structures are established within 100$\tau$, on the average. Their
further rearrangement requires much longer time. The $\beta_5$-strand
starts to interact with the $\beta_1$-strand at about 500$\tau$. Most
of the contacts that form the sandwich  are also established at this
time. The latest rearrangement, which appears to happen between
fragment L of the RT loop and strand $\beta_4$, occurs at about
1000$\tau$ and this  accomplishes the folding.

Among 100 of the trajectories
there are:
\begin{itemize}
\item 50 trajectories with scenario: \\
$\beta_3+\beta_4 \longrightarrow \beta_1+\beta_2
\longrightarrow \beta_1+\beta_5 \longrightarrow L+\beta_4$,
\item 25 trajectories with scenario: \\
$\beta_3+\beta_4 \longrightarrow \beta_1+\beta_2
\longrightarrow L+\beta_4 \longrightarrow \beta_1+\beta_5$,
\item 8 trajectories with scenario: \\
$\beta_1+\beta_2 \longrightarrow \beta_3+\beta_4
\longrightarrow \beta_1+\beta_5 \longrightarrow L+\beta_4$,
\item 7 trajectories with scenario: \\
$\beta_1+\beta_2 \longrightarrow \beta_3+\beta_4
\longrightarrow L+\beta_4 \longrightarrow \beta_1+\beta_5$,
\item 10 trajectories with other possible sequencings (not more
than 2 for each scenario).
\end{itemize}
One can notice that the sequencings among trajectories seem to be more
diversified than in the case of CI2.  The folding mechanism presented
by the average sequencing of contacts still dominates but only half of
the trajectories fold exactly according to this scenario.  However, in
75 out of 100 trajectories, the $\beta_3$-$\beta_4$ sheet is formed
earlier than the other segments. This appears to be in very good
agreement with numerous experimental studies
\cite{Grantcharova,spectrin,Martinez1,Martinez,Riddle,Plaxco2} which
reported that the distal loop hairpin is clearly present in the
transition state.  Notice also that in 90 out of 100 trajectories the
$\beta_1$-$\beta_2$ and $\beta_3$-$\beta_4$ sheets are formed before
the $\beta_1$-$\beta_5$ sheet and the $L$-$\beta_4$ contacts are
established. This is consistent with a recent $\Phi$-value analysis by
Martinez and Serrano \cite{Martinez}, which suggests that the folding
of SH3 seems to be composed of two folding subdomains. One of such
domain consists of the $3_{10}$-helix and the distal loop hairpin,
while the other corresponds to the RT loop.  

\begin{figure}
\epsfxsize=3.0in
\centerline{\epsffile{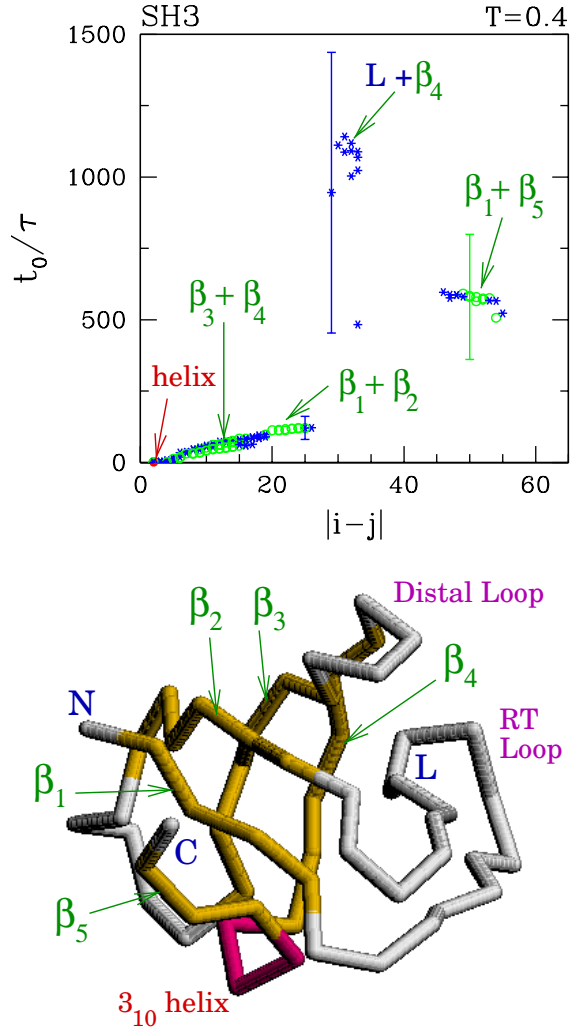}}
\caption{The same like Figure 5, but for SH3, at $T$=0.4.}
\end{figure}

Like for CI2 we have checked that increasing (by a factor of 5)
the strengths of the contacts in the hydrophobic core does not lead to
any significant change in the sequencing. For SH3, our chosen 
contacts that correspond to the hydrophobic core are
Leu86 - Ile111, Leu101 - Ile133, Ile111 - Glu121
and Glu121 - Ile133.
Experimentally, it has been shown \cite{Martinez1} that certain
mutations in the $\alpha$-spectrin SH3 can significantly increase both
thermodynamic stability and folding rates without affecting transition
states.  The structure of the transition state has also been shown to
be highly conserved among the SH3 homologies.
\cite{Martinez,Riddle,Plaxco2} This indicates that the folding
mechanism of SH3 depends primarily on the  native topology instead of
on the specific interactions.  Thus our simulation results with
increasing the hydrophobic interactions are consistent with those
experimental findings. 

\section{Conclusions}

In summary, we have studied three globular proteins in the Go-like
models with the Lennard-Jones potentials for the native contacts.  We
have delineated the native basins of the models by several (contact
and distance based) criteria which were found to be almost equivalent
in practice.  The models were found to be stable and well folding.  We
have found that there is a strong correlation between the time to form
a contact and its corresponding distance along the sequence.  This is
consistent with recent observation by Plaxco et al. \cite{Plaxco} on
the dependence of the folding rates on the contact order parameters of
the native states.  In general, the $\alpha$-helices which are
stabilized mainly by the local contacts appear much faster than the
$\beta$-sheets.  History of folding contains several gaps when folding
is hampered temporarily.  Usually, trajectories which evolve according
to the average scenario are the most frequent kind of possible
trajectories.  The folding scenarios are found to be in agreement with
studies on the structure of the transition states.

We also find that the average sequencing of the folding events does
not depend on the viscosity of the environment.  Furthermore, large
changes in the strength of certain contact energies or small changes
in the temperature affect the sequencing very little.  All this
indicates that the folding evolution depends primarily on the geometry
of the native conformation. 

We are grateful to Jayanth R. Banavar for useful discussions.  This
work was supported by Komitet Badan Naukowych (Poland; Grant Number
2P03B-025-13). Figures 1, 5, 8, 10 and 11 are prepared with the help
of the program RasMol version 2.6.

\end{document}